\documentclass[fleqn,twoside]{article}
\usepackage{espcrc2}

\usepackage{graphicx}
\usepackage[figuresright]{rotating}

\newcommand{\AmS}{{\protect\the\textfont2
  A\kern-.1667em\lower.5ex\hbox{M}\kern-.125emS}}

\title{Hadronic Multi-Particle Final State Measurements with CLAS at Jefferson Lab}
\author{W. K. Brooks\thanks{This work is supported in part by the U.S. Department of Energy, including
                DOE Contract No. DE-AC05-84ER40150.}\\*[0.25cm]
        Physics Division\\Thomas Jefferson National Accelerator Facility\\Newport News, Virginia, U.S.A.}
       
\begin{document}

\begin{abstract}
Precision measurements in the neutrino sector are becoming
increasingly feasible due to the development of relatively high-rate
experimental capabilities. These important developments command
renewed attention to the systematic corrections needed to interpret
the data. Hadronic multi-particle final state measurements made using
CLAS at Jefferson Lab, together with a broad theoretical effort that
links electro-nucleus and neutrino-nucleus data, will address this
problem, and will elucidate long-standing problems in intermediate
energy nuclear physics. This new work will ultimately enable precision
determinations of fundamental quantities such as the neutrino mixing
matrix elements in detailed studies of neutrino oscillations. 
\vspace{1pc}
\end{abstract}

\maketitle

\section{Introduction}

Neutrino physics has evolved over the past three decades from a niche
area with limited experimental exposure to a forefront field having
important applications to numerous fundamental studies. The impact of
the observation of neutrino oscillations has been profound, since it
is a clear example of physics that may be unknown to the Standard
Model. The modern revolution in astrophysics has been powerfully
shaped by neutrinos as their potential role in supernovae and dark
matter has clarified. These points of progress have stimulated
intensive improvements in neutrino experiments, such as high rate,
long-baseline experiments designed to characterize neutrino oscillations.

A concomitant of achieving high fluxes, however, is that 
the understanding of systematic uncertainties in the measurements must
be revisited. In particular, since most experiments use nuclear
targets as a partial means of attaining higher rates, a better
understanding of the lepto-nuclear response becomes
critical in interpreting statistically precise data from neutrino
detectors. Symbiotically, an improved understanding of leptonuclear
physics in the neutrino sector is likely to emerge, particularly in
combination with data from \emph{electro-nuclear} experiments.

Electron scattering experiments have exerted an important influence on
nuclear and particle physics for over half a century. From the
discovery in the 1950's that protons are not pointlike, to the
discovery of 'scaling' in deep inelastic scattering (DIS) that pointed
to the existence of quarks, to the precise mapping of form factors of
nuclei and nucleons, to the understanding of the spin structure of the 
proton, electron scattering has been the 'microscope of choice' for
understanding the structure of composite systems.

However, these achievements have been based on inclusive measurements,
detecting only the scattered electrons. Studies of other particles
emerging from the interactions has historically been badly hampered by
low rates relative to hadron beams, which have much higher cross
sections. The essential technical limitation on 
rates was the very low duty factor of accelerating cavities operating
at high gradient, due to thermal limitations; experimental rates were
therefore constrained by peak pulse intensities. The breakthrough
technology of superconducting accelerating structures has removed this
limitation, bringing the advent of accelerators with 100\% duty
cycle ('continuous wave', or 'cw'). The first institution to implement
a large-scale cw accelerator was the
Thomas Jefferson National Accelerator Facility (Jefferson Lab) in
Newport News, Virginia, USA, which houses the Continuous Electron Beam
Accelerator Facility (CEBAF)\cite{CEBAF}. CEBAF began operation in 1995.

CEBAF provides a 100 micron diameter electron beam of more than 100
microamperes with energies in the range 0.5-5.7 GeV to three
experimental areas simultaneously. Beam 
bunches arrive at the targets every two nanoseconds, permitting
high-luminosity coincidence measurements. Of the three experimental
areas, Hall A and Hall C provide multiple high-resolution magnetic
spectrometers with momentum acceptances ranging from 5\% to 20\% and angular 
acceptances ranging from 5 to 10 millisteradians, while Hall B houses a moderate resolution, large
acceptance spectrometer. This instrument, the CEBAF Large Acceptance
Spectrometer (CLAS) in Hall B\cite{CLAS}, is capable of detecting multiple
final-state charged and neutral hadrons over a wide range of angles
and momenta. In the following sections, a description of the
experimental capabilities of CLAS and its potential role in
interpreting data from neutrino facilities is presented.

\begin{figure}[h!tb]
\includegraphics[scale=0.5]{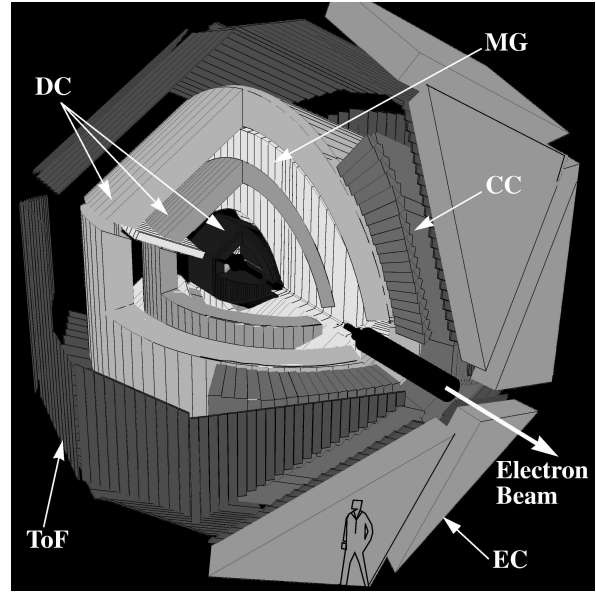}
\begin{center}
\end{center}
\vspace{-40pt}
\caption{A three-dimensional representation of the CEBAF Large
Acceptance Spectrometer, CLAS. The cutaway reveals a view
of the interior of the spectrometer. The five major subsystems are indicated
in the figure: the superconducting torus magnet (MG), the three regions of
drift chambers (DC), the time-of-flight counters (ToF), the Cerenkov
counters (CC), and the electromagnetic shower calorimeters (EC).} 
\label{fig:clas_3d}
\end{figure}

\section{Description of CLAS}

The primary instrument in Jefferson Lab's Hall B, CLAS was constructed
by the CLAS collaboration, an international alliance of 160
scientists from 39 academic institutions. Data taking began at the end of
1997. The scientific program centered around CLAS is exceptionally
broad in scope, including: 
\begin{itemize}
\item{a comprehensive investigation of the properties of excited
nucleons, with emphasis on searching for new resonances}
\item{an epic study of the neutron magnetic form factor}
\item{polarized structure functions of the proton and deuteron}
\item {deep virtual Compton scattering and its connection to
generalized parton distributions}
\item {quark/hadron propagation through nuclear systems}
\item{semi-inclusive deep inelastic scattering on nucleons and nuclei}
\item{exotic meson searches}
\item{elementary and nuclear hyperon production}
\item{two-nucleon correlations in light nuclei}
\end{itemize}

\begin{figure}[h!tb]
\begin{center}
\includegraphics[scale=0.44]{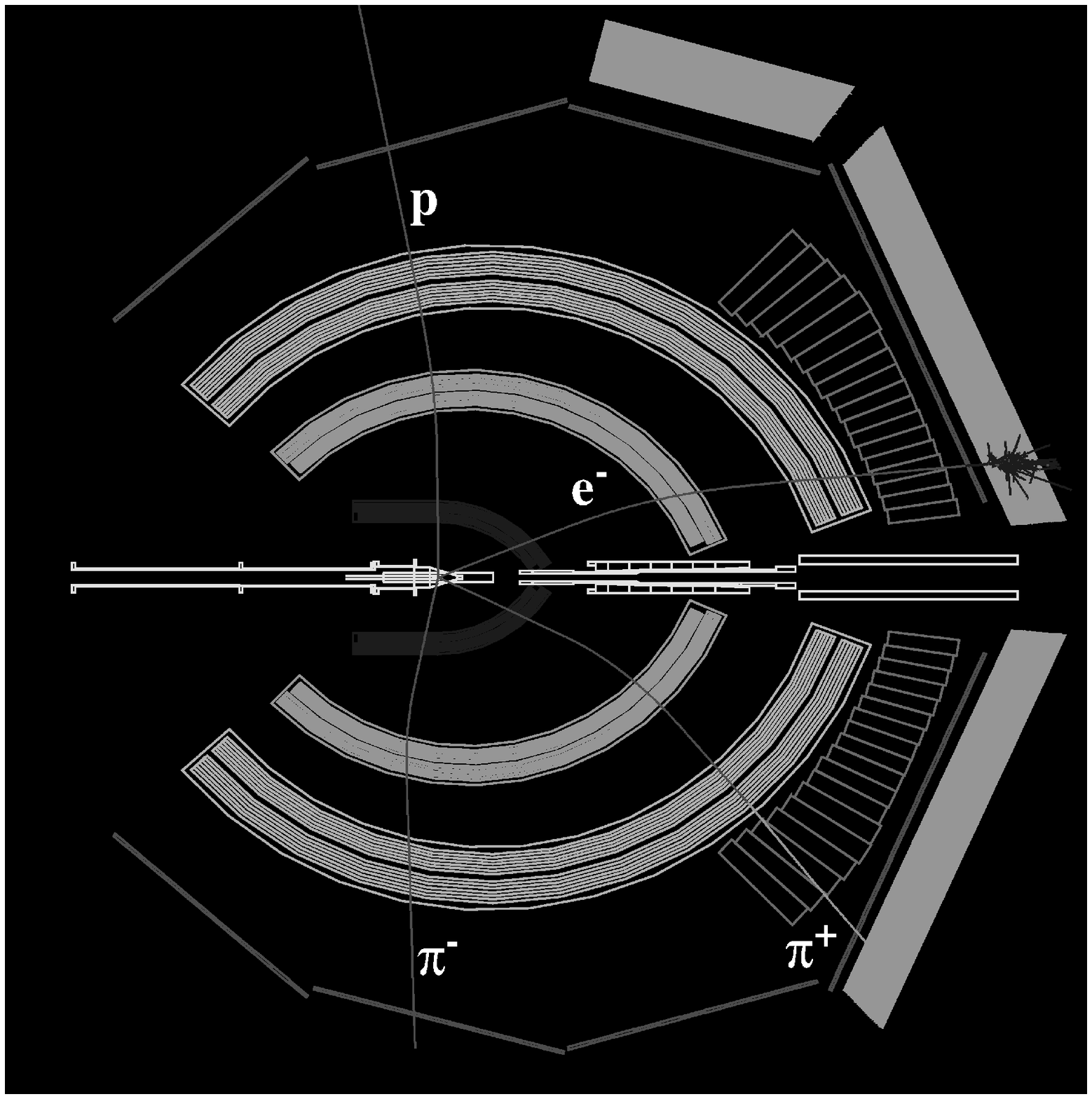}
\includegraphics[scale=0.38]{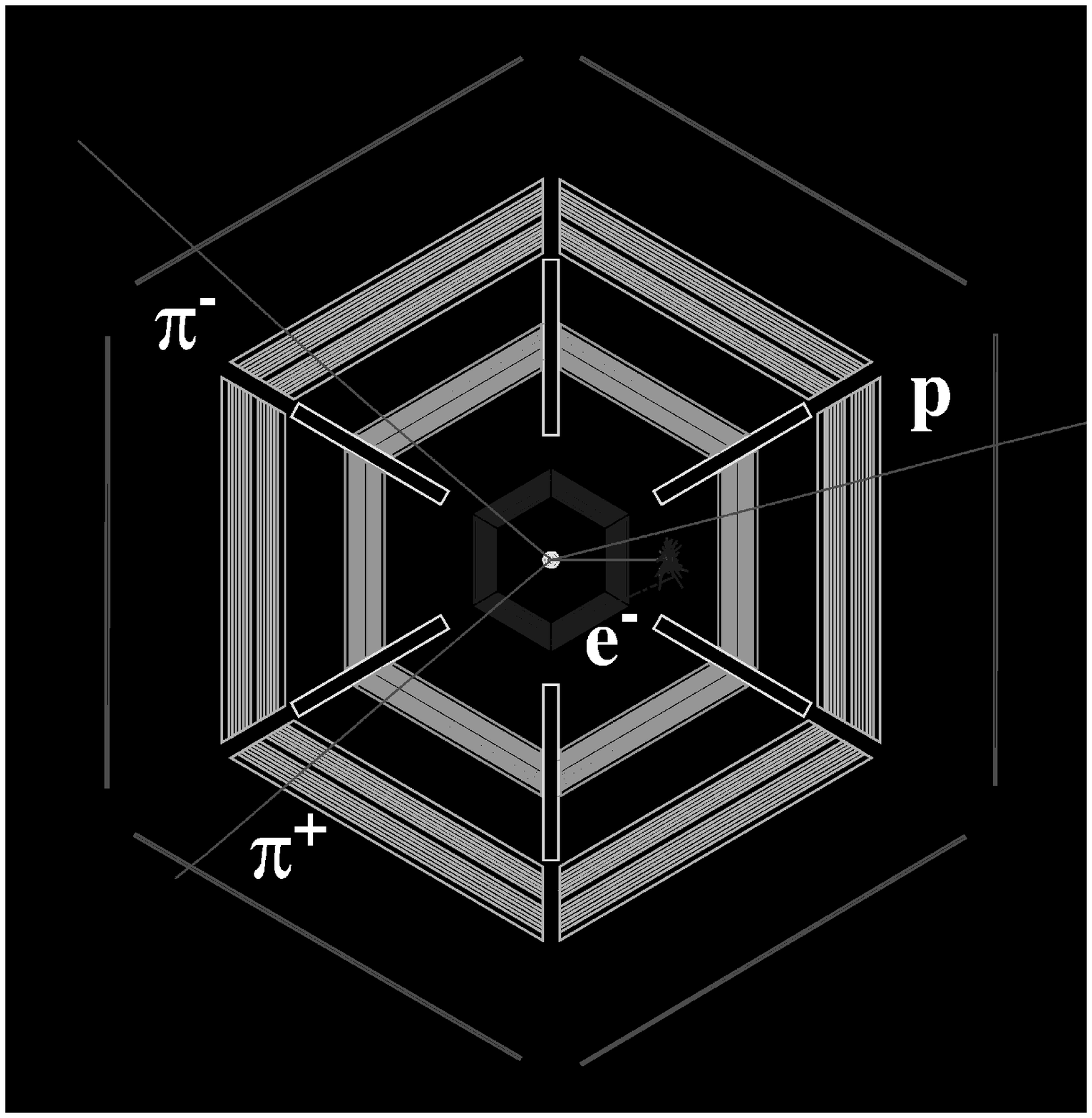}
\end{center}
\vspace{-30pt}
\caption{Two projected views of CLAS. The upper panel shows a
horizontal cross
section of CLAS containing the target region, viewed from
above. The lower panel shows a view along the beamline in a plane
containing the target. Charged particle tracks can be seen in these
views: with the nominal torus polarity, negatively charged particles
bend toward the beamline while positive particles bend away.} 
\label{fig:clas_views}
\end{figure}

\begin{figure}[h!tb]
\begin{center}
\includegraphics[scale=0.4]{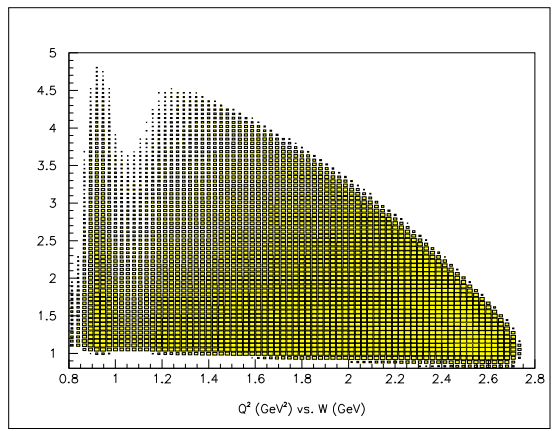}
\end{center}
\vspace{-30pt}
\caption{A plot of the electron variables from the CLAS E5 run. The
plot shows a sample of 50 million electrons acquired with a beam
energy of 4.232 GeV. The cutoff in $Q^2$ is a function of beam energy
and magnetic 
field settings; the lower practical limit is approximately 0.1 $GeV^2$.}
\label{fig:clas_q2vsw}
\end{figure}

The experimental requirements to satisfy this broad program include a
few common elements that mandate employing a large acceptance
spectrometer. These elements include the capability of measuring
final states with multiple, nearly uncorrelated particles; or
measurements that intrinsically require low beam current. Examples of the
latter include polarized target experiments, and photon beam experiments using
Bremsstrahlung photons that are energy-tagged by the associated electron.

The technical solution chosen to achieve large acceptance in an
electron beam environment was a toroidal magnetic field, where the
incident electron beam is directed along the toroidal axis of
symmetry. This field configuration offers several advantages over
solenoid or dipole fields: for instance, it has full acceptance over a wide range in
scattering angle, and is field-free at the target location, which is
critical for polarized target operation.
A three-dimensional view of CLAS is depicted in
Fig. \ref{fig:clas_3d}, where the cutaway reveals details of the inner detector
configuration. The field is produced by six kidney-shaped
superconducting magnet 
coils sheathed in six thin stainless steel cryostats. This choice
naturally segments the detector volume into six sectors. Three drift
chamber layers sample the charged particle trajectory before, during,
and after the bend in the magnetic field. In the forward direction,
electron identification is obtained at the trigger level by a
combination of gas Cerenkov counters and an electromagnetic shower
calorimeter, which subtend scattering angles from 8 to 45 degrees from
the nominal target location, while the drift chambers span the range 8
to 142 degrees in scattering angle. This angular range is matched by
an array of 5 centimeter thick scintillators which are optimized for
precision time measurements; for many of the charged hadrons of
interest, the time-of-flight technique can determine the hadron mass,
thus distinguishing charged pions, kaons, protons, and
deuterons. There are also calorimeters at larger 
angles in two sectors. In addition to electron and positron detection
and pion 
rejection, all calorimeters are used for neutron and neutral pion
detection. Two-dimensional 'slices' through the detector are shown in
Fig. \ref{fig:clas_views}. In the upper panel, the electron beam is
incident from the left; the trajectories of charged
particles may be seen through the drift chambers and magnetic field
volume. The electron track can also be seen to interact with the forward
electromagnetic shower calorimeter. In the lower panel of this figure,
the acceptance in the azimuthal angle for this four-track event is
evident. 

The field-free central region of the spectrometer offers great
flexibility in the selection of targets. Cryogenic liquid hydrogen,
deuterium, $^3$He and $^4$He have all been used, as well as solid
targets of carbon, iron, and lead. Cryogenic polarized
solid NH$_3$ and ND$_3$ targets with superconducting
Helmholtz coils have also been used for double polarization
measurements.

\begin{figure}[h!tb]
\begin{center}
\includegraphics[scale=0.4]{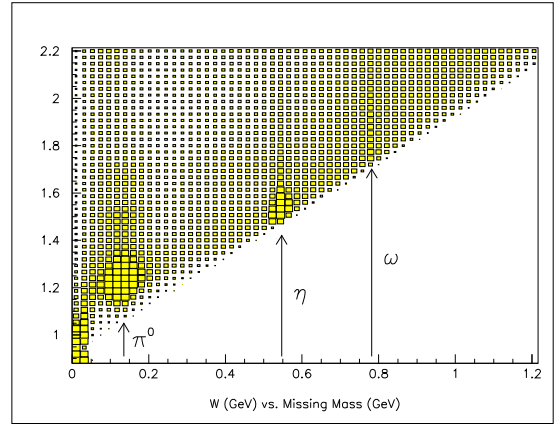}
\end{center}
\vspace{-30pt}
\caption{The invariant mass W vs. the missing mass of the unmeasured
particles in electron scattering from a hydrogen target, where one proton is detected in
coincidence with the electron. The coupling of many of the resonance
states to particular hadronic decay channels is visible by eye. }
\label{fig:clas_wvsmm}
\end{figure}

\section{Description of CLAS Data}

For electron beam experiments, the spectrometer is usually operated at
its 
design luminosity of 
$10^{34}~cm^{-2}s^{-1}$; this produces nominally a megahertz rate of
hadronic interactions and a gigahertz rate of M{\o}ller electrons. The
drift chambers are protected from the latter by a small toroidal
magnetic shield produced by a 
normal-conducting magnet located inside the innermost drift chamber. A
highly-configurable fast trigger preferentially selects events where
a scattered electron intercepts all three drift chambers, reducing
the accepted event rate to 3-4 kilohertz. More restrictive triggers are
used occasionally for specialized measurements. The data transfer rate
can exceed 20 megabytes per second, or approximately a terabyte per
day when operational efficiencies are taken into account. Depending on the
experiment configurations, from 10 to 30 billion triggers are
collected annually. 

In distinction to traditional electron scattering measurements, in
CLAS a wide range of angles and momenta are reconstructed
simultaneously at given beam energy. As a
result, continuous spectra are obtained for 4-momentum transfer squared ($Q^2$),
invariant mass of the hadronic system ($W$), and hadron kinematic
variables. An example is shown in Fig. \ref{fig:clas_q2vsw}, which illustrates the range
of $Q^2$ and $W$ obtained from a 4 GeV beam. 

\begin{figure*}[h!tb]
\begin{center}
\includegraphics[scale=0.5,angle=-90]{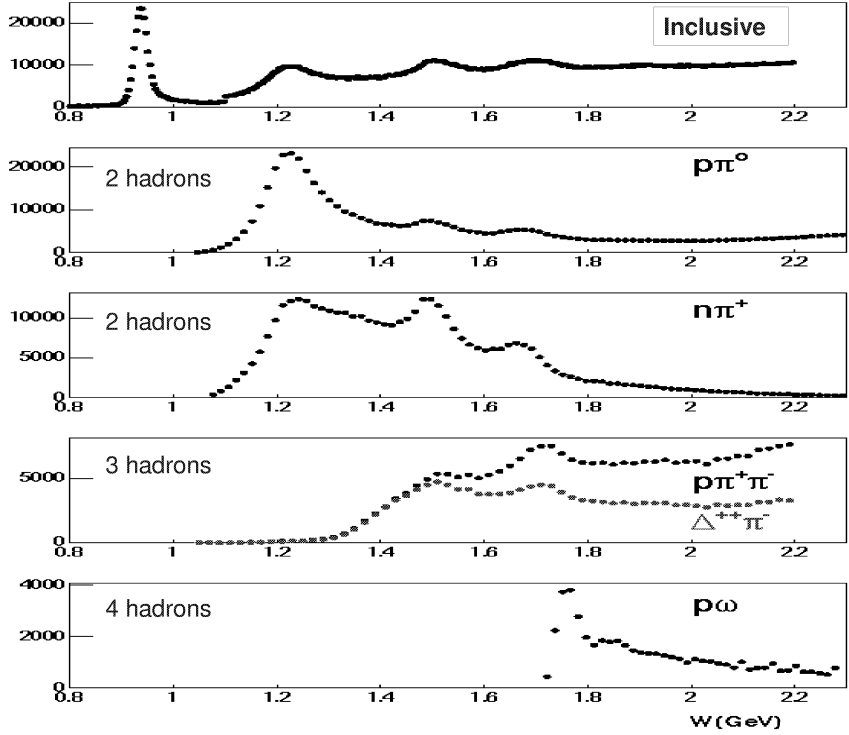}
\end{center}
\vspace{-30pt}
\caption{An example demonstrating the multi-hadron reconstruction capability
of CLAS. The top plot shows the inclusive W spectrum; the successive
plots show the W spectra when two, three, and four hadrons are
detected. In the lower four plots, one of the hadrons is inferred by the
missing mass technique while the other hadrons are detected directly.}
\label{fig:clas_multihad}
\end{figure*}

A typical event reconstruction begins by identifying the scattered
electron for a given 
event using information from the Cerenkov detector and the
electromagnetic shower calorimeter. Subsequently, hadrons are
identified by calculating the time-of-flight (TOF) of other charged tracks
in the event and deriving their mass from the TOF and the measured
momentum. Neutral particles can be measured directly, or can also be
inferred using the missing mass technique. As an example, in the reaction
$e+p~\rightarrow~e'+p+X$ where a scattered electron and proton are measured,
the mass of the unmeasured system X can be calculated. With
the typical resolution of 10-20 MeV, missing hadrons can easily be
identified. An illustration of this can be seen in Fig. \ref{fig:clas_wvsmm}, in which W
is plotted against missing mass. Here the decay of resonances with
masses in the range 1.2-2.2 GeV into $\pi^0$, $\eta$, and $\omega$
mesons is easily identified. A further illustration of this technique
applied to successively more complex events is shown in Fig. \ref{fig:clas_multihad}, where
up to four hadrons are measured in addition to the scattered electron.

\section{Neutrino Connection}

Now that the basic phenomenon of neutrino oscillations has been
established, a program of measurements is needed for its
characterization. 

For instance, determination of the mass differences and
elements of the mixing matrix can be carried out with a combination of
high-statistics, 
accelerator-based long-baseline experiments; such experiments are
planned in the U. S. and Japan. The maximal probability of
observing neutrino oscillations for the relevant experiments occurs
for neutrino energies ranging from 0.5 - 3.0 GeV.

Neutrino and electron interactions with hadronic matter in this energy
range typically produce multi-hadron final states via complex 
interaction mechanisms. The elementary production mechanisms on
protons and neutrons include elastic scattering, resonance production,
and deep inelastic scattering. These mechanisms are modulated in
lepto\emph{nuclear} interactions by Fermi motion, Pauli blocking, coherent
pion production, hadron absorption and propagation through nuclei,
nuclear form factors, and hadron formation length effects. While some
of these topics have been thoroughly studied in the electron sector,
much less is known from neutrino measurements. Because neutrino
detectors are of necessity based on simple, integrating measurement
schemes, such as Cerenkov light emission in water, the precise details
of the final state are unknown and must be simulated. The complexity
of the neutrino-nuclear response, coupled to integrating detection
technologies, translates into uncertainties in the interpretation of
neutrino data.

However, electron scattering and neutrino scattering are closely
connected. The differences in the lepton tensor are trivial,
and much is known about the relationships between the hadronic 
currents interacting with the two types of probes. A systematic
theoretical treatment that describes hadron production in
electro-nuclear interactions can in turn be used to predict hadron
production in neutrino-nuclear interactions, beginning with the
elementary production, and continuing with nuclear targets. Because
CLAS offers multi-hadron final state measurements for few-GeV
electrons that can constrain such a theoretical effort, a new and
unique opportunity for a cross-disciplinary study of leptonuclear
physics now exists for the first time.

A full program of this type would involve developing or adapting
models to describe the elementary reactions on protons and
neutrons. The models should be designed so that they can describe both
electron beam and neutrino beam interactions. These 
could be tested 
against the large body of existing CLAS data for hydrogen and
deuterium targets. This data has been acquired for a number of beam
energies and experimental conditions; elastic, resonance, and DIS data
are essentially always taken together in the same dataset, although
the high W cutoff varies slightly depending on trigger
thresholds. While there already exist models for such interactions,
the requirements here are somewhat different than the typical purpose
for developing descriptions of the data, and will therefore involve
different choices. For instance, the specification of the higher
resonance contributions may be adequate at a more phenomenological
level than would be true of, e.g., a search for small contributions
from missing resonances, a typical motivation for CLAS 
measurements. At the same time, attaining a good description of the
data over a wider range in W may be more critical for neutrino studies
than, for instance, studying the properties of a particular resonance
in exhaustive detail, another typical motivation of CLAS
measurements. Above the two-pion threshold, the multi-hadron
production mechanisms for resonances may have to be treated with a
more phenomenological approach because of the complexity involved, but
the many-fold differential cross sections can be constrained well by
testing against CLAS data, even for two and three pion production. 

Once the models for elementary interactions are constructed and
tested, they can be turned to predict the interactions with nuclear
systems. Here a host of new effects must be considered, as were listed
above. At low energies and low momentum transfers, Pauli blocking,
coherent pion production, and nuclear form factors will measurably
alter the observed hadron yields. In the DIS regime, the formation
lengths of hadrons will come into play; the determination of these
formation lengths is a subject of one CLAS program involving nuclear
targets\cite{QPPROPOSAL}\cite{QPWRITEUP}. The effects of hadron
absorption and propagation 
through nuclei, as well as Fermi motion, have a strong impact on the
data at all relevant energies. Existing CLAS data on carbon and iron
targets, as well as $^2H$, $^3He$, and $^4He$, can be used to study
the A dependence of these effects; to access kinematic regions not yet
measured in CLAS, new data can be acquired. A particularly important
effect is the final state interaction influencing the reaction
products in the nucleus; since this large effect is
nearly identical for neutrino beams and electron beams, it can be
predicted with confidence for neutrinos once the electron beam data is
well-described. Needless to say, there are excellent prospects, not
only for studying neutrino-nucleus interactions, but also for
improving our understanding of the nuclear physics involved at a more
fundamental level. This cross-disciplinary, symbiotic motivation
holds the promise to involve a wider pool of talent and expertise in
these compelling studies.

\section{Conclusions}

The advent of cw electron-beam accelerators, together
with advances in detector and magnet technology, now permit
measurements of multi-hadron final states with high statistical
precision in many-fold differential cross sections. The combination of
electron beam data from CLAS and a broad-based 
theoretical effort connecting these data to neutrino scattering holds
great promise for advances in both neutrino science and in nuclear
physics. This new work will enable precision determinations of
fundamental quantities such as the neutrino mixing matrix elements in
detailed studies of neutrino oscillations.

\end{document}